\documentclass[12pt,preprint]{aastex}

\renewcommand{\b}[1]{\boldsymbol{#1}}

\newcommand{\unit}[1]{\,{\rm #1}}
\newcommand{\cm}{\unit{cm}}
\newcommand{\G}{\unit{G}}
\newcommand{\kG}{\unit{kG}}
\newcommand{\rpm}{\unit{rpm}}
\newcommand{\s}{\unit{s}}

\newcommand{\Rm}{Re_{\rm m}}

\providecommand{\boldsymbol}[1]{\mbox{\boldmath $#1$}}

\providecommand{\tabularnewline}{\\}

\shorttitle{2D convective instability simulations}
\shortauthors{W. Liu \emph{et al.}}

\begin{document}

\title{Noise-Sustained Convective Instability in a Magnetized Taylor-Couette Flow}

\author{Wei Liu\altaffilmark{1}}

\altaffiltext{1}{Current address: Theoretical Division, Los Alamos National Laboratory, Los Alamos, NM, USA 87545}

\affil{Center for Magnetic Self-Organization in Laboratory and Astrophysical Plasma, Princeton Plasma Physics Laboratory, Princeton, NJ, USA 08543}
\email{wliu@lanl.gov}





\begin{abstract}
The helical magnetorotational instability of the magnetized Taylor-Couette flow is studied numerically in a finite cylinder. A distant upstream insulating boundary is shown to stabilize the convective instability entirely while reducing the growth rate of the absolute instability. The reduction is less severe with larger height. After modeling the boundary conditions properly, the wave patterns observed in the experiment turn out to be a noise-sustained convective instability. After the source of the noise resulted from unstable Ekman and Stewartson layers is switched off, a slowly-decaying inertial oscillation is observed in the simulation. We reach the conclusion that the experiments completed to date have not yet reached the regime of absolute instability.

\end{abstract}

\section{Introduction}
The magnetorotational instability (MRI) is probably the main source of
turbulence and accretion in sufficiently ionized astrophysical disks
\citep{bh98}. Due to  this crucial role in astrophysics, substantial efforts have been spent worldwide to observe MRI in a laboratory setting \citep{jgk01,gj02,npc02,sisan04,vils06}, but MRI has never been conclusively demonstrated in the laboratory. 

\begin{figure}[!htp]
\begin{center}
\epsscale{0.4}
\plotone{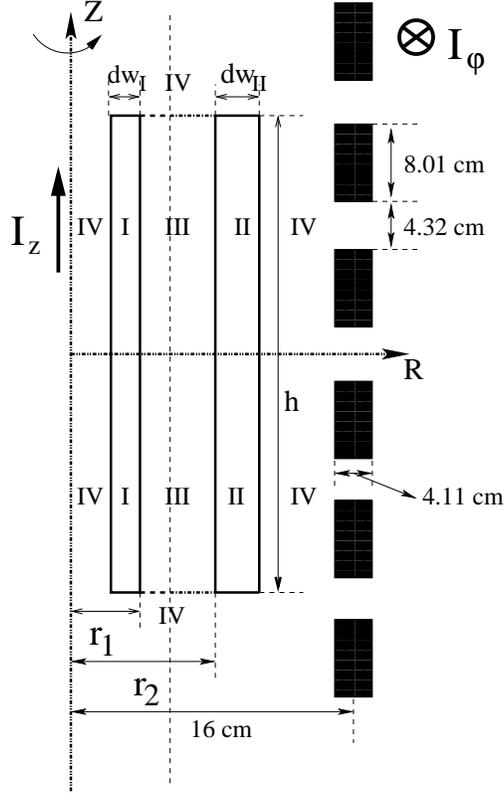}
\caption{
  Computational domain for simulations of PROMISE II experiment. Region
  (I): Inner copper cylinder, angular velocity $\Omega_{1}$.
  (II): outer copper cylinder, 
  $\Omega_{2}$. (III): liquid gallium; (IV):
  vacuum. Thick dashed line: insulating inner ring, corotating with the inner cylinder. Thick dash-dot line: insulating outer ring, corotating with the outer cylinder. The junction of these two rings lies at 40\% of the gap ($D=r_2-r_1$) between the inner and outer cylinder \citep{sj07}. Dimensions:
  $r_{1}=4.0\unit{cm}$; $r_{2}=8.0\unit{cm}$; $h=40.0\unit{cm}$;
  $d_{wI}=1.0\unit{cm}$; $d_{wII}=1.5\unit{cm}$;
  $\Omega_{1}/2\pi=3.6\unit{rpm}$; $\Omega_{2}/2\pi=0.936\unit{rpm}$. Note that $\mu=\Omega_2/\Omega_1=0.26$, rather than $\mu=0.27$ used in previous work \citep{sgg06,rhsgg06,lgj07,sgg07}.
  The exact configuration of the toroidal coils being unavailable to us,
  six coils (black rectangles) with dimensions as shown were used,
  with $67$ turns in the two coils nearest the midplane and $72$ in the rest.
  Currents $I_\varphi$ were adjusted to reproduce the reported 
  Hartmann numbers $Ha\equiv B_{z}^{0}r_{1}/\sqrt{\rho\mu_{0}\eta\nu}$.
\label{boundary} }

\end{center}    
\end{figure}

Most experiments have been done in cylindrical geometry with a background flow that approximates the ideal Couette rotating profile:
\begin{equation}
\label{couette}
\Omega=a+b/r^{2}
\end{equation}
where $a=(\Omega_{2}r_{2}^{2}-\Omega_{1}r_{1}^{2})/(r_{2}^{2}-r_{1}^{2})$
and $b=r_{1}^{2}r_{2}^{2}(\Omega_{1}-\Omega_{2})/(r_{2}^{2}-r_{1}^{2})$, $\Omega_1$ and $\Omega_2$ are the rotation speed of the inner and outer cylinder and $r_1$ and $r_2$ are the radius of the inner and outer cylinder, respectively (see Fig.~\ref{boundary}).
For axially periodic or infinite magnetized Taylor-Couette flow, MRI-like modes have been shown theoretically to grow at much reduced magnetic Reynolds number $\Rm\equiv\Omega_1r_1(r_2-r_1)/\eta$ and Lundquist number $S\equiv V_{A,0}r_1(r_2-r_1)/\eta$ in the presence of a combination of axial and current-free toroidal field %
\begin{equation}
\label{eq:backgroundfield}
\b{B}^{0}=B_z^{0}\left(\b{e}_z+\beta r_1/r \b{e}_\varphi\right)
\end{equation}
than the standard MRI (SMRI) with purely axial magnetic field \citep{hr05,rhss05}. Here the cylindrical coordinates $(r,\varphi,z)$ are used. $B_z^{0}$ and
$\beta$ are constants. The Alfv\'en speed is defined as $V_{A,0}\equiv B_z^{0}/\sqrt{4\pi\rho}$. $\eta$ and $\rho$ are the magnetic diffusivity and density of the fluid, respectively  (see Fig.~\ref{boundary}).
 
The Potsdam ROssendorf Magnetic Instability Experiment (PROMISE) group claimed to have observed this kind of helical MRI (HMRI) experimentally \citep{sgg06,rhsgg06,sgg07}. However we have shown that the wave pattern observed in PROMISE is not a global instability, but rather a transient disturbance somehow excited by the Ekman circulation and then transiently amplified as it propagates along the background axial Poynting flux with nonzero group and phase velocities, but is then absorbed once it reaches the jet formed at midheight between two neighboring Ekman cells \citep{lgj07}. PROMISE group have accordingly updated the experimental facility to PROMISE II to allow for two split rings at both endcaps: the inner ring attached to the inner cylinder and outer ring attached to the outer cylinder. If the width of the inner ring is chosen appropriately $\sim0.4(r_2-r_1)$, the magnetized Ekman circulation could be significantly reduced, therefore removing one of the possible disturbance sources, \emph{i.e.} the unsteady jet \citep{sj07}. 

As with other examples in the literatires, such as drifting dynamo waves \citep{tpk98, ptk00}, it is of vital importance to distinguish absolute instability from convective instability in a traveling wave experiment like PROMISE. It is also an essential ingredient of the threshold prediction for the Riga dynamo \citep{ggglps08}. For a traveling wave the positivity of the growth rate implies only an amplification of the perturbation as it moves downstream. In one case, despite the movement of the wave packet, the perturbation increases without limit in the course of time at any point fixed in space; this kind of instability with respect to any infinitesimal perturbations will be called \emph{absolute instability}. In the other case, the packet is carried away so swiftly that at any point fixed in space the perturbation tends to zero as $t\rightarrow\infty$; this kind will be called \emph{convective instability} \citep{ll87} (see the details of \S\ref{packet}). For PROMISE II, it appears that under the experimental conditions the second kind occurs. A recent preprint also highlights the importance of the distinction between absolute and convective instabilities in the context of the HMRI \citep{pg08}.

In a Taylor-Couette experiment bounded by insulating endcaps, \citet{tpk98} have pointed out that without any external disturbances except a small initial disturbance needed as a seed for the instability, the distant upstream insulating boundary acts as an ``absorbing" boundary while the characteristics of the downstream endcap is unimportant. Due to this absorption the convective unstable state cannot be sustained by a uniform driving force, therefore this unstable mode eventually decays \citep{tpk98}. This driving force is not the noise mentioned before, but the power to drive the instability, which in the usual Taylor-Couette experiments can be quantified by the magnetic Reynolds number $\Rm$. This conclusion has been rigorously demonstrated in the very resistive limit in \S II.C of \citet{lghj06} using a perturbative approach and \S II.D of \citet{lghj06} using a modifed WKB analysis, showing that the insulating endcap entirely stabilizes the HMRI mode, which is a convective unstable mode given the parameters of the PROMISE experiment. 

The absorbing boundary is essential to the development, regardless of how distant it may be. The larger height only defers the time when we have to wait for the boundary-induced dissipation to dominate \citep{tpk98}. On the other hand, if $\Rm$ exceeds a higher threshold $Re_{\rm m,f}$, the driving force of the system overcome the dissipation and a globally unstable mode appears \citep{tpk98}. Therefore in a bounded system the unstable mode appears at $Re_{\rm m,f}$ rather than $Re_{\rm m,c}$, where $Re_{\rm m,c}$ is the critical magnetic Reynolds number for the onset of the convective unstable mode without the ``absorbing" boundary. \citet{tpk98} has showed that in the presence of an ``absorbing" boundary and large $h$, a global unstable mode appears when
\[
\Rm\geqslant Re_{\rm m,f}\equiv Re_{\rm m,a}+O(h^{-2})\,,
\]
where $Re_{\rm m,a}$ is the critical magnetic Reynolds number corresponding to the onset of the absolute instability without the ``absorbing" boundary. 

This has raised a big obstacle for people to observe absolutely unstable HMRI in the laboratory. The advantage of HMRI itself, \emph{i.e.}, unstable with a low critical Reynolds number ($3$ orders lower than the SMRI) conflicts with the necessarily high threshold of the onset of an absolute HMRI mode, \emph{i.e.}, excited at a reasonably high critical magnetic Reynolds number, thus high Reynolds number, which would result in much more severe end-effects than people had expected. Moreover the fact that the critical Lundquist number must usually increase together with the magnetic Reynolds number and high ratio of toroidal-to-poloidal magnetic field requirement ($\beta>1$) would even worsen the situation. 

We also find that by nonlinear numerical simulation the insulating endcap reduces the growth rate of the absolute instability somewhat. The higher the height $h$ is, the less the growth rate is reduced (Table~\ref{reduce}). 

In a typical experiment, the experiment is, however, highly likely affected by small external noise either from a physical cause or experimental imperfection such as the misalignment of the cylinders. If the system is convectively unstable, \emph{i.e.}, disturbances grow as they move downstream, noise would sustain structures in the system even if no global mode is unstable \citep{dr87,ptk00}. In the present paper, we show by numerical simulations that the perturbations from the unstable magnetized residual Ekman layer and Stewartson layer at the upper endcap would play the role of ``noise" generator, though this perturbation level is reduced with increasing axial magnetic field \citep{lw08a}. What is observed in PROMISE II turns out to be a noise-sustained convective traveling wave, not the absolute unstable mode.

This paper is organized as follows: 
\S\ref{packet} presents the wave packet analysis in a unbounded cylinder, which is the basis of the following sections. 
We report the nonlinear simulation results with \emph{partially} conducting boundary conditions of PROMISE II experiment in \S\ref{noise}. The final conclusions and implications to the HMRI experiments are given in \S\ref{discussion}.

\section{Wave Packet Analysis in an Unbounded Cylinder}\label{packet}
Assuming a cylinder of infinite height $h$, $k_{z}$ is a continuous variable.
Let the gap width be fixed and finite, so $k_{r}\cong\pi/(r_{2}-r_{1})$. We define the total wavenumber $K=\sqrt{k_{r}^{2}+k_{z}^{2}}$ and the growth rate $\gamma$.

Since the fast growing mode is the dominant mode, here we focus on waves with vertical wavenumber $k_z$ close to that of the fastest growing mode, $k_z^0$. The range of values of $k_z$ lies near the point for which $\gamma(k_z)$ is a maximum, \emph{i.e.} $d\gamma/d k_z=0$ at $k_z=k_z^0$ [as seen from Fig.~\ref{fig:global} (a)]. Let  a slight perturbation occurs near the middle of the flow $(z\sim0)$ in the format of a wave packet as follows:
\begin{equation}
\label{wavepacket}
B_{r}(z,t=0)=b_{0}\exp\left(-\frac{z^{2}}{2L^{2}}\right)\exp(ik_{z}^{0}z)\, ,
\end{equation}
where we have used the envelope $\exp(-z^{2}/2L^{2})$ to confine the perturbation around the central part of the cylinder, where $L\sim O(h)$. In the course of time, the components for which $\gamma(k_z)>0$ will be amplified, while the remainder will be damped. The amplified wave packet thus formed will also be carried downstream with a velocity equal to the group velocity $d\omega/dk_z$ of the packet, where $\omega=\mathbb{R}\omega+i\gamma$ and $\mathbb{R}\omega$ is the real part of the frequency; since we are now considering waves whose wave numbers lies in a small range near the point where $d\gamma/dk_z=0$, the quantity
\begin{equation}
\label{group_velocity}
V_g=d\omega/dk_z \cong d(\mathbb{R}\omega)dk_z
\end{equation} 
is real, and is therefore the actual velocity of propagation of the packet. This downstream displacement of the perturbations is very important, and causes the complications of absolute instability \emph{v.s.} convective instability.

We can approximate the dispersion relation like (Fig.~\ref{fig:global}):
\begin{eqnarray}
\label{approx1}
\mathbb{R}\omega=\mathbb{R}\omega(k_{z})=\kappa\frac{k_{z}}{K}\,; \\
\label{approx2}
\gamma=\gamma(k_{z})=\gamma^{0}-\frac{\sigma}{2}(k_{z}-k_{z}^{0})^{2}\,,
\end{eqnarray}
in which $\kappa^{2}=\frac{1}{r^{3}}\frac{d}{dr}(r^{2}\Omega)^{2}=4(1+Ro)\Omega^{2}$ and $Ro\equiv1/2d\ln \Omega/d\ln r=a/\Omega-1$ is the Rossby number. We know $\gamma=0$ when $k_{z}=0$. Thus
$\sigma=2\gamma^{0}/k_{z}^{02}$. And in order to simplify the derivation, we assume $K\approx constant $ from now on (though this is not a good approximation, we can get some insightful results from this simple approximation). From Eq.~\ref{approx1}, we get
$V_g=\kappa/K$.

At later time $t>0$
\begin{eqnarray}
\label{eqn5}
\widetilde{B_{r}}(k_{z},t)&=\widetilde{B_{r}}(k_{z},0)\exp(\gamma(k_{z})t+i\mathbb{R}\omega(k_{z})t) \nonumber\\
                                           &=\widetilde{B_{r}}(k_{z},0)\exp\{[\gamma^{0}-\frac{\sigma}{2}(k_{z}-k_{z}^{0})^{2}]t+i\kappa\frac{k_{z}}{K}t\}\,,
\end{eqnarray}
if we define $D=\sqrt{L^{2}+\sigma t}$, the result can be expressed as:
\begin{equation}
\label{eqn20}
B_{r}(z,t)=b_{0}\frac{L}{D}\exp(\gamma^{0}t)\exp\left[-\frac{(z+V_gt)^{2}}{2D^{2}}\right]\exp[ik_{z}^{0}(z+V_gt)]\,.
\end{equation}
In Eq.~\ref{eqn20}, As $t\rightarrow0$, Eq.~\ref{eqn20} can be simplified as:
\begin{equation}
\label{eqn21}
B_{r}(z,t)=b_{0}\exp(\gamma^{0}t)\exp[ik_{z}^{0}(z+V_gt)]\,,
\end{equation}
which is a ``transient" growing phase.
As $t\rightarrow\infty$, 
\begin{equation}
\label{convective}
B_{r}(z,t)=b_{0}\frac{L}{\sqrt{\sigma t}}\exp\left[\left(\gamma^{0}-\frac{V_g^2}{2\sigma}\right)t\right]\exp[ik_{z}^{0}(z+V_gt)]
\end{equation}

Obviously, If $\gamma^0<\gamma_{\rm a}=V_g^2/2\sigma$, we will get \emph{convective instability}, that is, it starts with a transiently growing phase (Eq.~\ref{eqn21}), followed by a phase asymptotically decaying to zero (Eq.~\ref{convective}).
If $\gamma^0>\gamma_{\rm a}$, we will get \emph{absolute instability}.

\section{Noise-Sustained Convective Instability in PROMISE II Experiment}\label{noise}
In order to reduce the undesirable effects induced by the endcaps and also the accompanying hydromagnetic asymmetries, \citet{sj07} have proposed to split both endcaps into two rings which are attached to both cylinders and found that if the width of the inner ring is chosen to be $0.4D$ (see Fig.~\ref{boundary}), where $D=r_2-r_1$ is the gap between the inner and outer cylinder, the magnetic energy in term of $b_{\varphi}$, where $b_{\varphi}$ is the perturbed azimuthal magnetic field, is minimized. Therefore the magnetized Ekman circulation is significantly reduced, leading to a satisfactory ideal Couette state (Eq.~\ref{couette}) in the bulk flow. PROMISE has been accordingly updated to PROMISE II adopting this idea.

\begin{figure}[!htp]
\begin{center}
\epsscale{0.8}
\plotone{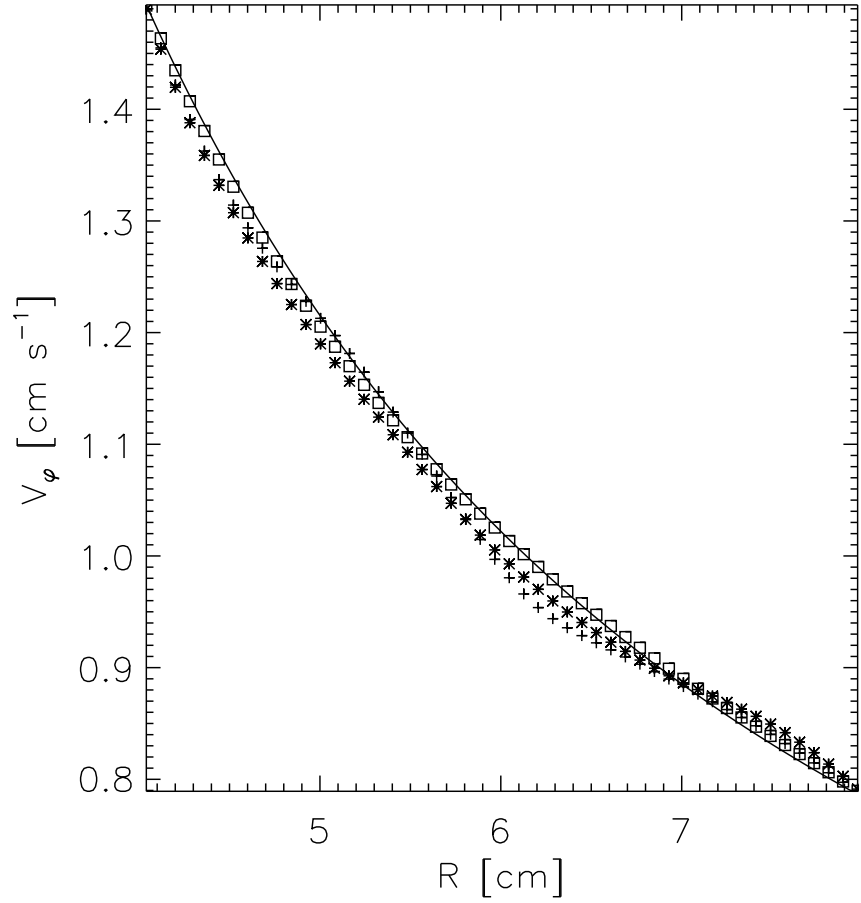}
\caption{~Azimuthal velocity \emph{v.s.} Radius $r$.  $Re=1775$, $\beta=3.81$ and $I_{\varphi}=75\unit{A}$. Solid line,~ideal Couette state; $+,~1.31\;\unit{cm}$;
$*$,~$2.72\;\unit{cm}$; $\Box,~13.95\;\unit{cm}$. \label{noekman} }

\end{center}    
\end{figure}

While we have confirmed their conclusions (Fig.~\ref{noekman}) (Please note that in \citet{sj07}, this conclusion is derived with $\beta=0$, \emph{i.e.}, no background toroidal magnetic field, while our simulation results show that this conclusion is also valid with nonzero $\beta$), here we report nonlinear simulations with the ZEUS-MP
2.0 code \citep{hnf06}, which is a 
time-explicit, compressible, astrophysical ideal MHD parallel 3D code,
to which we have added viscosity, resistivity (with subcycling to reduce
the cost of the induction equation), and \emph{partially} conducting boundary conditions \citep{lgj07}, for axisymmetric flows in cylindrical coordinates
$(r,\varphi,z)$. It has been demonstrated that the finite conductivity ($\eta_{\rm Cu}= 1.335\times10^{2}\unit{cm^{2}s^{-1}}$) and thickness of the
copper vessel are important,
and this noticeably improves agreement with the measurements compared
to previous much simplified boundary condition \citep{lgj07}. Please note that in this paper $\mu=\Omega_2/\Omega_1=0.26$, rather than $\mu=0.27$ reported in previous work.
The parameters of PROMISE II as reported in or inferred from
\citet{sgg08} are used: gallium density
$\rho=6.35\unit{g\;cm^{-3}}$, magnetic diffusivity $\eta=2.43\times10^{3}\unit{cm^{2}\;s^{-1}}$, magnetic Prandtl
number $Pr_{\rm m}\equiv\nu/\eta=1.40\times10^{-6}$; Reynolds number $Re\equiv
\Omega_{1}r_{1}(r_{2}-r_{1})/\nu=1775$; axial current
$I_{z}=6000\unit{A}$; toroidal-coil currents $I_\varphi=0, 50, 75,
120\unit{A}$; and dimensions as in Fig.~\ref{boundary}. 

\begin{figure}[!htp]
\epsscale{1.0}
\plottwo{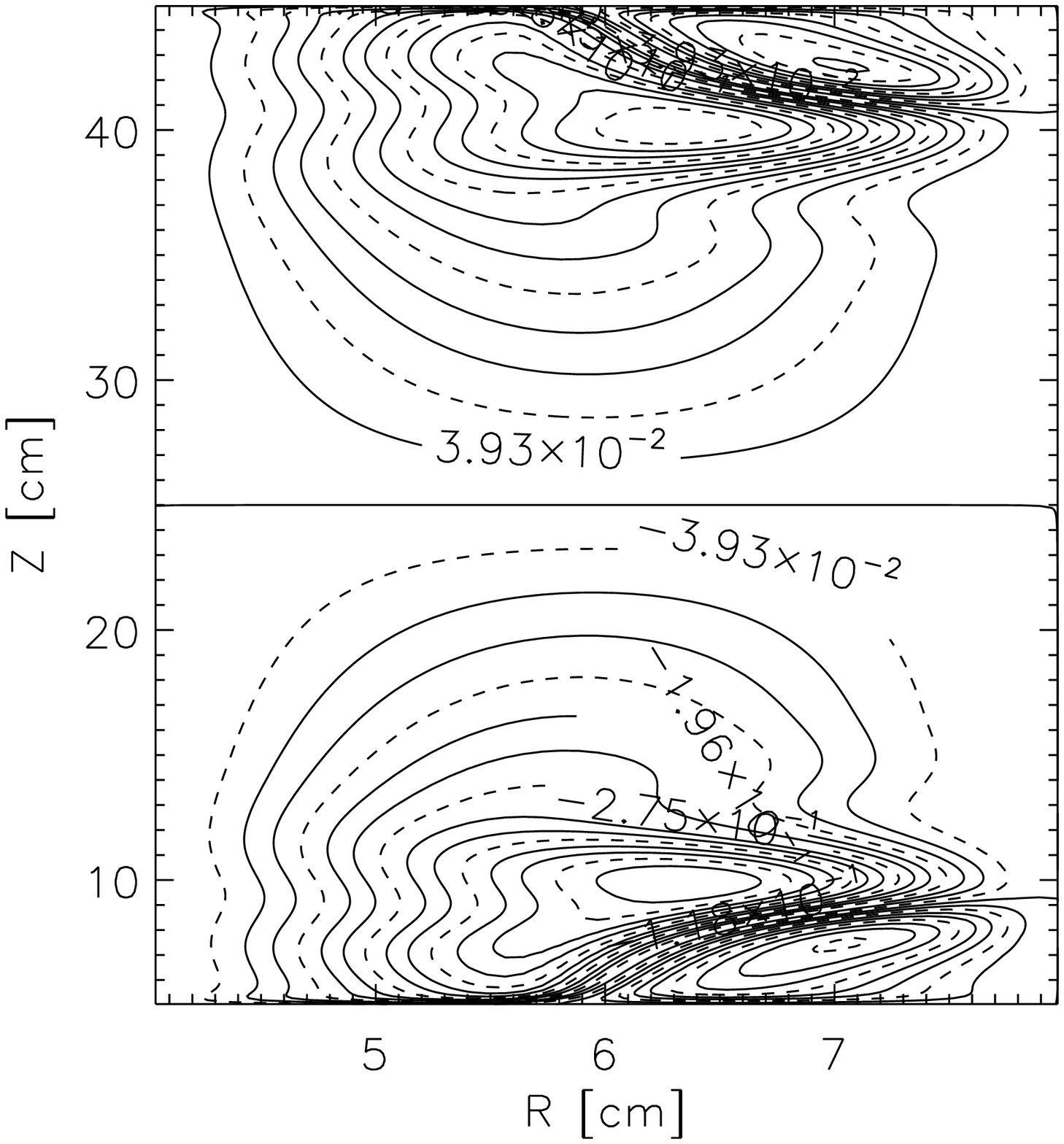}{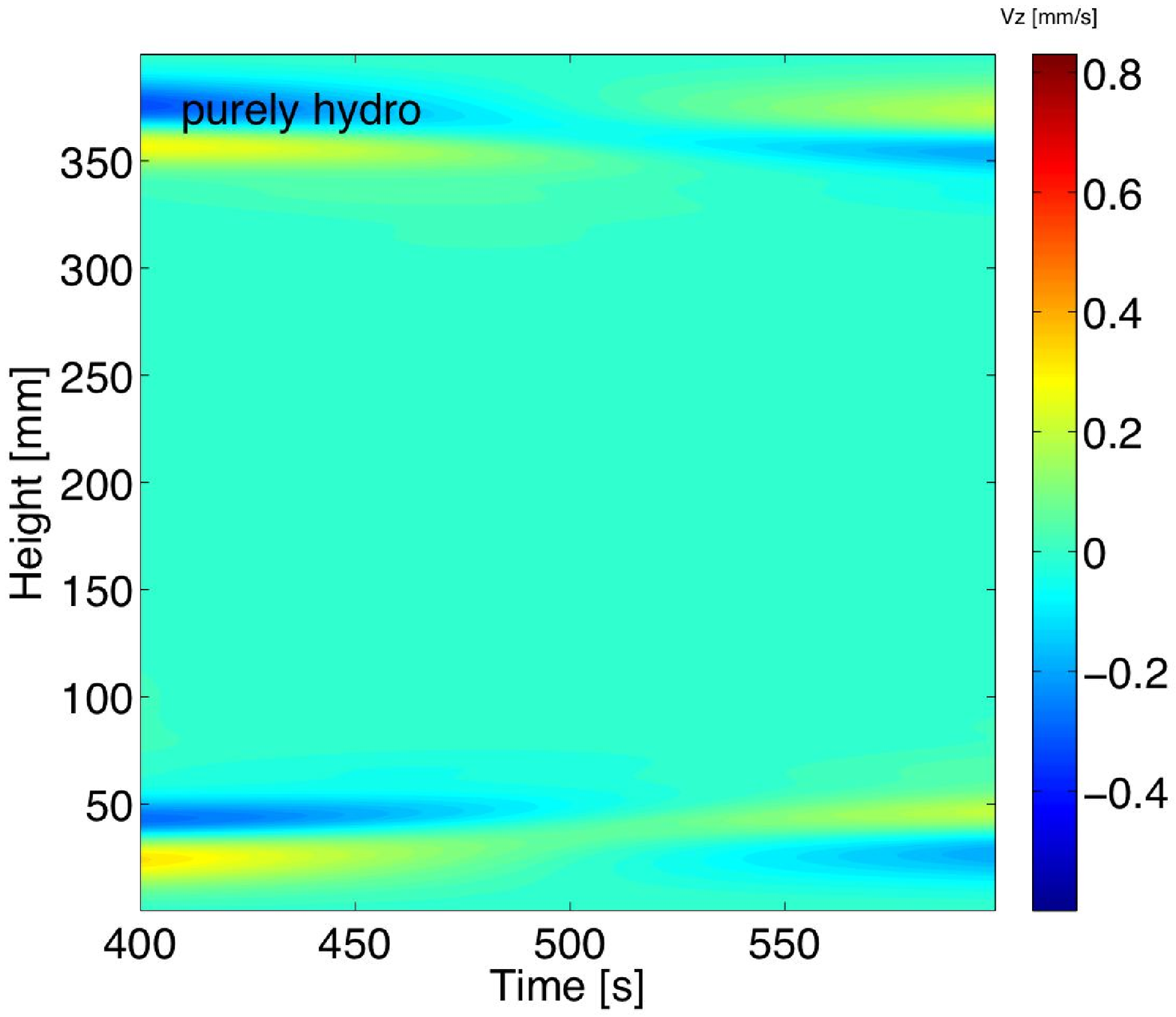}
\caption{\label{hydro} Purely hydrodynamic (unmagnetized) simulations. \emph{Left}: Time-averaged poloidal flow stream function $\Psi$; \emph{Right}:~(color) Axial velocities $[\unit{mm\;s^{-1}}]$ versus time and
  depth sampled at $r=6.5\;\unit{cm}$, for
  the parameters of the PROMISE II experiment without any magnetic field. Note height increases upward from the bottom endcap. 
  No-slip velocity boundary conditions are imposed at the rigidly rotating
  endcaps. The steady part of the resulting Ekman circulation is
  suppressed in right panel by subtracting the time average at each height. }
\end{figure}

\begin{figure}[!htp]
\begin{center}
\epsscale{1.0}
\plotone{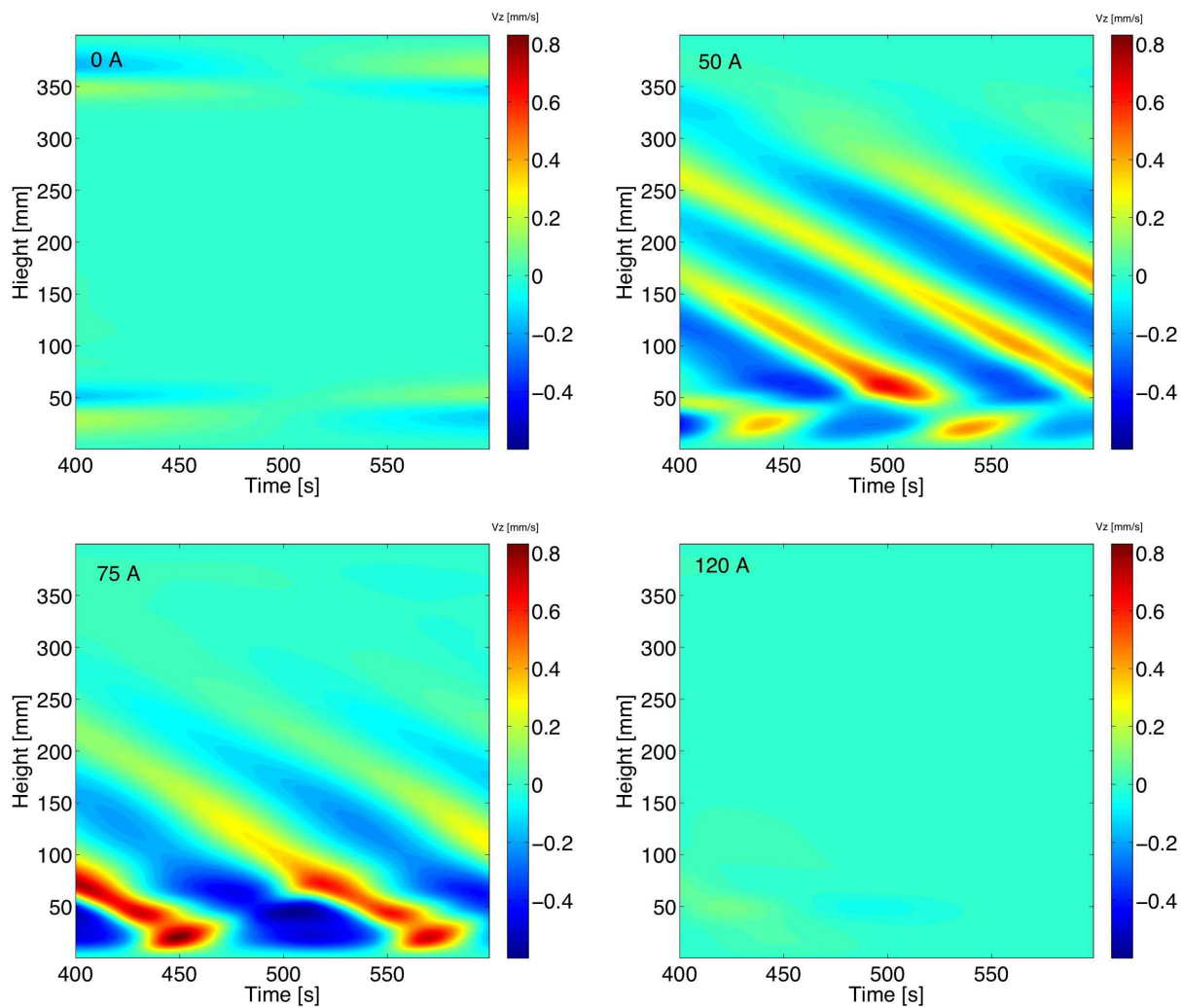}
\caption{
  (color). Axial velocities $[\unit{mm\;s^{-1}}]$ versus time and
  depth sampled at $r=6.5\cm$, for
  the parameters of the PROMISE II experiment 
  with toroidal currents $I_\varphi$ as marked. 
  No-slip velocity boundary conditions are imposed at the rigidly rotating
  endcaps. The steady part of the resulting Ekman circulation is
  suppressed in these plots by subtracting the time average at each height.
  The waves appear to be
  absorbed near the bottom endcap.
  \label{wave} }

\end{center}    
\end{figure}

For comparison, we start with purely hydrodynamic (unmagnetized) simulations (Fig.~\ref{hydro}). From Fig.~\ref{hydro} (a), after splitting the endcaps into two rings, the two big Ekman cells are divided into four smaller cells and localized near the endcaps. Compared to the simulation results of PROMISE \citep{lgj07}, there is not an flapping ``jet" near the mid-plane as in the usual Ekman circulations. This removes the possible noise from this unsteadiness. However from Fig.~\ref{hydro} (b), there are some perturbations near both endcaps, which supply the possible sources of noise in the system. These perturbations are resulted from unstable Ekman layer and Stewartson layer \citep{lw08a}. The magnitude of this noise is around $\pm0.2\unit{mm\;s^{-1}}$. As we will see later (Fig.~\ref{wave}), this unsteadiness is reduced by increasing axial magnetic field \citep{gp71,lw08a}.
 
Figure.~\ref{wave} displays vertical velocities
near the outer cylinder in simulations corresponding
to the experimental runs of \citet{sgg08} for several values
of the toroidal current, $I_\varphi$.
A wave pattern very similar to that in the
experimental data  \citep{sgg08} is seen. Since now there is no jet, the traveling wave is propagating to the bottom endcap and absorbed there while in the old PROMISE experiment, the traveling wave disappears at the jet \citep{lgj07}. 
We also notice that the perturbation near the upper endcap weakens with strong axial magnetic field. This could be explained by a more stable magnetized residual Ekman layer and Stewartson layer \citep{lw08a}. Both the weakening of the noise sources and disappearance of the amplifying mechanism leads to a rather steady state with $I_{\varphi}=120\unit{A}$.

\begin{figure}[!htp]
\begin{center}
\epsscale{0.8}
\plotone{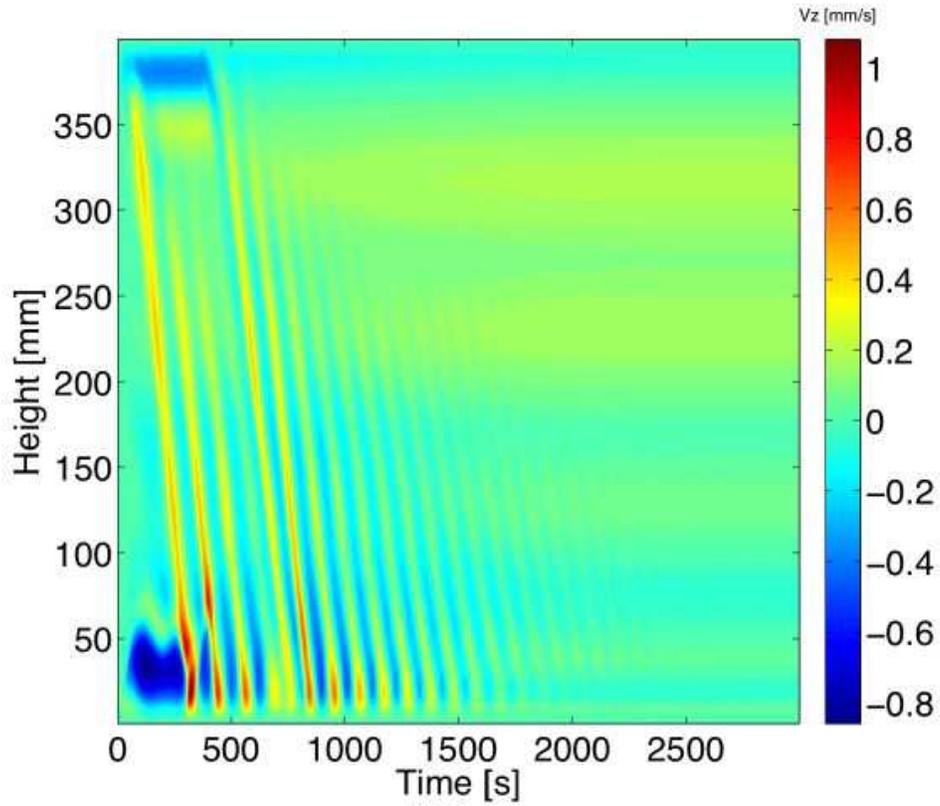}
\caption{
(color). An extended version of the case
 $I_\varphi=75\unit{A}$ shown in Fig.~\ref{wave} but without
 subtraction of the time average. After $t=360\unit{s}$, 
  the no-slip boundary condition at both endcaps is switched to an
  ideal Couette profile (Eq.~\ref{couette}). A slowly decayed inertial oscillation is resulted.   
  \label{decay} }

\end{center}    
\end{figure}

It is highly possible that there is much noise in the real experiment due to some experimental imperfection such as misalignment and in the numerical simulation such as numerical noise. Also the noise could result from physical causes such as the unsteady Ekman layer or Stewartson layer. These noises would cause a noise-sustained convective instability in the system as in \citet{ptk00}. The continuous impulse from the noise sources would have the system always in the state of ``transiently growing" phase (Eq.~\ref{eqn21}). This results in similar wave patterns as the ones from the primary instability without noise, which are observed in PROMISE and PROMISE II experiments and simulations \citep{lgj07}. The noise-induced wave pattern is always susceptible to noise-induced disruption as discussed by \citet{dr87}. That is exactly what we found here and in \citet{lgj07}. We can see this point more clearly by following \citet{lgj07}: performing
a simulation that begins with the experimental boundary conditions until
the traveling waves are well established, and then switches abruptly
to ideal-Couette endcaps (Fig.~\ref{decay}). After the switch, the traveling waves disappear after one axial propagation time and slowly decaying inertial oscillations (asymptotically to zero) result. The main difference in results between \citet{lgj07} and the present simulation are: (1) there is no jet, thus the traveling waves are absorbed near the bottom endcap both before and after the switch; (2) there is no change of wave speed associated with the switch since the background state does not change much before and after the switch. We reach the conclusion that even after the endcaps are split into two rings as in PROMISE II, the wave patterns observed in the experiment are not global instability, but rather noised-sustained convective instability. The similarity between these ``inertial-oscillation-induced wave" after the switch and the earlier noise-sustained ``MRI waves" in the simulation or ``MRI-type waves" observed in the various versions of PROMISE stems from the physical nature of HMRI that HMRI is a weakly destabilized inertial oscillation \citep{lghj06}. More importantly, this similarity supports our conclusion in another aspect: the frequency and wave number selection mechanism for a noise-sustained structure is determined by a linear mechanism, thus resembling the properties of the primary instability. 


\section{Discussion}\label{discussion}
In this paper, nonlinear simulations of the helical magnetorotational instability in a magnetized Taylor-Couette flow are performed. The geometry mimics PROMISE II experiment with endcaps split into two rings. The partially conducting boundary condition introduced in \citet{lgj07} is used. The waves patters change with applied magnetic field as in the experiment. However via numerical tests, we find that the wave patterns observed in PROMISE II experiment are not due to a global instability, but rather a noise-sustained convective instability. 

The importance of the distinction between absolute and convective instability in a bounded system with broken reflection symmetry is discussed. The addition of the toroidal magnetic field breaks the axial symmetry of the system. In such cases, the effects of distant upstream insulating boundaries on the absolute instability differs remarkably from the ones on the convective instability. The insulating endcap would only reduce the growth rate of the absolute instability, but would stabilize the convective instability entirely, however distant it may be. For the absolute instability, the more distant insulating endcap would less reduce the growth rate, while for the convective instability the more distant endcap would only have the system wait longer for the dissipation due to the ``absorption" boundary to dominate. These discoveries cast great obstacles for people to observe the helical magnetorotational instability in the laboratory: An absolute HMRI is needed to observe the global unstable mode in the experiment.

\begin{figure}[!htp]
\epsscale{1.0}
\plotone{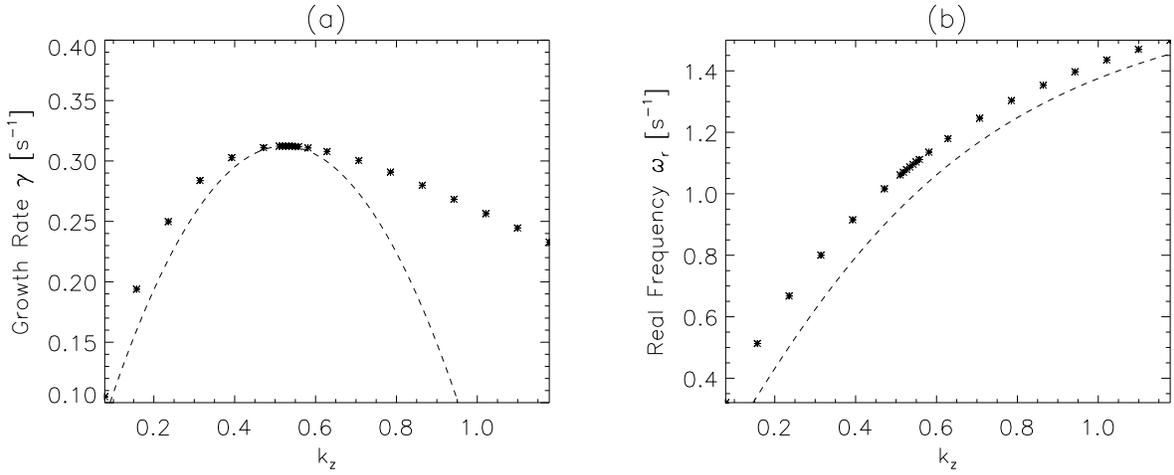}
\caption{\label{fig:global} (a) Growth Rate $\gamma$; (b) Real Frequency $\mathbb{R}\omega$.  *~ Linear Calculation, -~ Approximation by Eq.~\ref{approx1} and Eq.~\ref{approx2}. $r_{1}=4.0\cm$, $r_{2}=8.0\cm$,
$\Omega_{1}=101.25\rpm$, $\Omega_{2}=26.325\rpm$, $B_{z}=220.5\G$,
$\beta=4.0$; the material properties are based on
gallium: $\eta=2.43\times10^{3}\cm^2\s^{-1}$, $\nu=3.4\times10^{-3}\cm^2\s^{-1}$ and $\rho=6.35\unit{g\;cm^{-3}}$. The calculations are performed using a code \citep{gj02} adapted to allow for a helical field. Vertical periodicity is assumed, but the radial equations are
solved directly by finite differences with perfectly conducting
boundary conditions (\S II.B of \citet{lghj06}).}
\end{figure}

Unfortunately it is not easy to derive the critical magnetic Reynolds number $Re_{\rm m,f}$ of the absolute HMRI analytically in a bounded system. However we can get a rough estimate of $Re_{\rm m,a}$, \emph{i.e.}, the critical magnetic Reynolds number of the absolute HMRI in an unbounded system, by wave packet analysis (\S\ref{packet}) and the approximate dispersion relation from Fig.~\ref{fig:global}. From Fig.~\ref{fig:global}, we derive the group velocity $V_g\sim1.08\cm\s^{-1}$, $\gamma^0\sim0.31\s^{-1}$, $k_z^0\sim0.52\cm^{-1}$ and $\sigma\sim2.29\cm^2\s^{-1}$. Therefore $\gamma^0-V_g^2/2\sigma\sim0.05\s^{-1}>0$, which corresponds to an absolute HMRI instability with $Re_{\rm m, a}\sim0.07$. We therefore conjecture that $Re_{\rm m,f}\equiv Re_{\rm m,a}+O(h^{-2})\gtrsim0.07$ in PROMISE II. The critical magnetic Reynolds number is somehow one order of magnitude lower than the standard MRI, but still requires Reynolds number $Re\sim10^5$. Therefore we need to rotate the cylinder typically with more than one hundred $\unit{rpm}$. Such rotation rates are of course achievable, however with such a Reynolds number the advantage of HMRI with much lower Reynolds number, thus much lower end-effects, is not so great as people had expected.  Moreover in most HMRI unstable modes $\beta>1$ is preferred, this suggests a toroidal magnetic field typically at $\sim1,000\G$, which requires axial currents $>10^4\unit{A}$ inside the inner cylinder. This is a big engineering challenge in itself. The technical constraints prevent us to try to find the threshold by nonlinear numerical simulations such as: (1)~the current code can not afford large Reynolds number $(\sim10^{5})$, which is required for HMRI to enter the absolutely unstable regime; (2)~from the global linear calculation, the HMRI mode is stabilized if an artificially low Reynolds number like $\sim10^3$, which could be afforded by the current code, is employed.  

The non-axisymmetric $m=1$ modes are observed in the experiments \citep{sgg06,rhsgg06,sgg07}. Unfortunately since the simulations presented in this paper are all axisymmtric, the possibility to study this important mode is excluded. The extension of the current work to 3D will be the subject of the future study. Ruediger et al have already done some excellent work on this issue and found that given PROMISE parameters the nonaxisymmetric HMRI modes are always harder to be excited than the symmetric mode \citep{rhss05,rs08}.

\acknowledgments
The author would like to thank Jeremy Goodman and Hantao Ji for their very inspiring discussion and constructive comments. The author would also like to thank James Stone for the advice on the ZEUS
code, Stephen Jardin for the advice to implement fully insulating boundary conditions and Frank Stefani for pointing out the distinction between the convective instability and absolute instability in a bounded Taylor-Couette experiment at 2007 APS-DPP annual meeting. 
This work was supported by the US Department of Energy, NASA
under grants ATP03-0084-0106 and APRA04-0000-0152, the
National Science Foundation under grant AST-0205903.


\clearpage

\begin{table}[!htp]
\begin{center}
\begin{tabular}{c|c|c|c|c}

\hline 
$\Omega_1 h/V_{A,0}$&
 20.3 &
 40.6 &
 81.2 &
 periodic \tabularnewline
\hline
\hline 
Growth Rate $\gamma\;\unit{s^{-1}}$&
0.27&
0.58&
0.82&
1.06\tabularnewline
\hline

\end{tabular}
\caption{\label{reduce} Influence of the height $h$ upon the growth rate $\gamma$ of the absolute instability in a bounded cylinder. $r_{1}=7.1\cm$, $r_{2}=20.3\cm$,
$\Omega_{1}=400\rpm$, $\Omega_{2}=53.3\rpm$, $B_{z}=500\G$,
$B_\theta(r_1)=1\kG$, the height $h=27.9\;\unit{cm}$, $55.8\;\unit{cm}$ and $111.6\;\unit{cm}$; the material properties are based on
gallium: $\eta\approx2000\cm^2\s^{-1}$ and $\rho\approx6\unit{g\;cm^{-3}}$, which give $\Rm=2$ and $S=2.7$, no explicit viscosity present. The simulations are performed using a modified version of the astrophysical code ZEUS2D \citep{sn921,sn922,lgj06,lw08b}. The boundary conditions adopt the one introduced in \S II.D of \citet{lghj06}. Please note that no-slip boundary conditions are employed on all applicable boundaries and ideal Couette state (Eq.~\ref{couette}) is enforced at both endcaps in order to remove the Ekman circulation and possible disturbances induced by this boundary layer effect. The one labeled  ``periodic" uses vertically periodic boundary conditions with periodicity length $h=27.9\;\unit{cm}$.}
\end{center}
\end{table}

%


\clearpage

\end{document}